\documentstyle[prl,aps]{revtex}

\def\la{\langle}
\def\ra{\rangle}

\def\lb{\lbrack}
\def\rb{\rbrack}

\def\Tr{\mbox{Tr}}
\def\tr{\mbox{tr}}

\def\be{\begin{eqnarray}}
\def\ee{\end{eqnarray}}

\def\g5{\gamma_5}
\def\hg5{\hat{\gamma_5}}
\def\hP{\hat{P}}
\def\hw{\hat{w}}
\def\C{{\cal C}}
\def\hC{\hat{\cal C}}
\def\U{{\cal U}}
\def\A{{\cal A}}
\def\B{{\cal B}}

\def\V{{\cal V}}

\def\wB{\widetilde{B}}

\def\wPsi{\widetilde{\Psi}}

\def\D{{\cal D}}

\def\index{\mbox{index}\,}
\def\hG{\hat{\Gamma}}
\def\wA{\tilde{A}}

 \def\Slash#1{
  \begin{picture}(5,6)(0,0)
  \put(-.7,-1.2){\line(5,6)6}
  \end{picture}
  \kern-.8em#1}
 \def\slash#1{
  \begin{picture}(5,6)(0,0)
  \put(-1.5,-1.7){\line(5,6)5}
  \end{picture}
  \kern-.8em#1}

\def\sd{\Slash D}

\begin{document}
 
\draft

\title{Index of a family of lattice Dirac operators and its relation
to the non-abelian anomaly on the lattice}

\author{David H. Adams}

\address{Math. dept. and Centre for the Subatomic Structure of Matter, 
University of Adelaide,
Adelaide, S.A. 5005, Australia. E-mail: dadams@maths.adelaide.edu.au}

\date{29 October 2000}

\maketitle

\begin{abstract}
In the continuum, a topological obstruction to the vanishing of the
non-abelian anomaly in 2n dimensions is given by the index of a certain
Dirac operator in 2n+2 dimensions, or equivalently, the index of a 
2-parameter family of Dirac operators in 2n dimensions. 
In this paper an analogous result is 
derived for chiral fermions on the lattice in the Overlap formulation.
This involves deriving an Index Theorem for a family of lattice Dirac
operators satisfying the Ginsparg--Wilson relation. The index density
is proportional to L\"uscher's topological field in 2n+2 dimensions.

\end{abstract}


\widetext

The Atiyah--Singer Index Theorem, both for single operators \cite{AS(single)}
and families of operators \cite{AS(families)}, has been of major importance
in the modern development of Quantum Field Theory.
For example, the Index Theorem for single operators gives topological
insight into the non-vanishing of the axial anomaly \cite{Fuji(cont)},
provides the basis for a resolution of the U(1) problem \cite{tHooft} and
determines the dimension of the instanton moduli space (used in semiclassical
investigations of Yang--Mills theory; see, e.g., \cite{Schwarz}), while
the Families Index Theorem reveals topological obstructions to the vanishing
of gauge anomalies, thereby providing constraints on allowable
theories \cite{AG,AS(anomaly),Nash(book)}.

When attempting to get a well-defined non-perturbative formulation of
QFT's (in particular, gauge theories) one successful and well-established
approach has been to formulate the theory on a spacetime lattice 
\cite{Wilson}. Therefore, carrying over the Index Theorem, both for single
operators and families, to the lattice is an interesting and important
problem. For a long time this did not seem possible though, due to the
fermion doubling problem and the resulting need for acceptable lattice
Dirac operators to break chiral symmetry \cite{NN}.
In the traditional formulations, at best
only a remnant of the Index Theorem is retained on the lattice \cite{Smit}.
However, the situation has changed quite dramatically in recent years
with the advent of the Overlap formulation \cite{overlap} and the discovery 
\cite{Neu(PLB1+2),Hasenfratz,Laliena}
of acceptable lattice Dirac operators satisfying the Ginsparg--Wilson 
relation \cite{GW} 
\be
D\g5+\g5{}D=aD\g5{}D\qquad\quad\mbox{($a$=lattice spacing)}
\label{1}
\ee
Such operators
have exactly chiral zero-modes (since $D\psi=0\;\Rightarrow\;D(\g5\psi)
=(aD\g5{}D-\g5{}D)\psi=0$), which allows to define
$\index{}D\equiv\Tr(\g5|_{\ker{}D})$ \cite{Laliena}. There is a
``Lattice Index Theorem'' \cite{Laliena,L(PLB),Fuji(NPB)}
\be
\index{}D=-\frac{a}{2}\Tr(\g5{}D)=a^4\sum_xq(x)\,.
\label{2}
\ee
where
\be
q(x)=-\frac{a}{2}\tr(\g5{}D(x,x))
\label{3}
\ee
is the index density. For $SU(N)$ gauge fields on the Euclidean
2n-dimensional torus, $\index{}D$ and $q(x)$ reduce to the continuum
index and density in the classical continuum limit \cite{DA(Chiral99)},
at least when $D$ is the overlap Dirac operator \cite{Neu(PLB1+2)}.
(Earlier results in this direction were obtained in
\cite{Laliena,Fuji(NPB),KY+DA+Suzuki+ChiuH}.
When $D$ is the overlap Dirac operator the right-hand side of (\ref{2})
has a spectral flow interpretation which had previously been used
as a definition of lattice topological charge in \cite{overlap}.)
Furthermore, although it is not invariant
under the usual chiral transformations, the fermion action
$S=a^4\sum_x\bar{\psi}(x)D\psi(x)$ exhibits an exact lattice-deformed 
version of chiral symmetry \cite{L(PLB)} 
(which was implicit in the overlap formalism): $\delta{}S=0$ for 
$\delta\psi=\hg5\psi\,$, $\delta\bar{\psi}=\bar{\psi}\g5$ where
\be
\hg5=\g5(1-aD)
\label{7}
\ee
An easy consequence of (\ref{1}) is $\hg5^2=1$. Furthermore, after 
supplementing (\ref{1}) with the $\g5-$hermiticity condition
\be
D^*=\g5{}D\g5
\label{7a}
\ee
we have $\hg5^*=\hg5$. Thus $\hg5$ can be viewed as a lattice-deformed
chirality matrix. The axial anomaly for the lattice-deformed chiral symmetry
transformation above can be determined from the 
corresponding change in the fermion
measure to be $\A(x)=-ia\tr(\g5{}D(x,x))=2iq(x)$ \cite{L(PLB)}. 
This is completely analogous to the relation between
axial anomaly and index density in the continuum \cite{Fuji(cont)}.

Having seen that there is an exact Lattice Index Theorem for lattice Dirac
operators satisfying the GW relation, and that the index and its density
are related to the axial anomaly in precisely the same way as in the 
continuum, it is natural to ask if there is also a Lattice Index Theorem 
for {\it families} of such operators such that the families index is
related to gauge anomalies (or more precisely, to obstructions
to the vanishing of these anomalies) in the same way as in the continuum. 
In this paper we show that this is indeed the case:
We derive an Index Theorem ((\ref{100}) below) for a family of such lattice 
Dirac operators, parameterised by a 2-sphere in the orbit space of $SU(N)$ 
lattice gauge fields on the 2n-dimensional torus $T^{2n}$.
This is the prototype for a more general Lattice Families Index Theorem 
which is currently under development \cite{DA(inprogress)}. 
We find that this index is related to an obstruction to
gauge-invariance in precisely the same way as in the continuum setting,
where it was previously studied by Alvarez-Gaum\'e and Ginsparg \cite{AG}.
Furthermore, the index density is found to be proportional to L\"uscher's
topological field $q(x,y_1,y_2)$ in 2n+2 dimensions \cite{L2} (given by
(\ref{a10}) below). This provides a natural origin for $q(x,y_1,y_2)$
in the lattice theory. (It was introduced in an ad hoc manner in
\cite{L2}.) This is of interest and potential use in connection with
L\"uscher's approach towards achieving gauge-invariance in nonabelian
lattice chiral gauge theory: a local gauge anomaly-free formulation exists
if and only if the local cohomology class represented by $q(x,y_2,y_2)$
is trivial \cite{L2}.

In the continuum, the modulus of the chiral
determinant $\det(i\sd_+^A)$ (suitably regularised as in \cite{AG}) is
gauge-invariant, but anomalies may arise in the phase.
Consider a family $\phi_{\theta}$ of gauge transformations
parameterised by $\theta\in{}S^1$. 
Each $\phi_{\theta}$ is a map $T^{2n}\to{}SU(N)$ 
(we assume for simplicity that the fermion is in the fundamental 
representation).
The action of $\phi_{\theta}$ on $A$ determines a 
circle-family $\{A^{\theta}\}_{\theta\in{}S^1}$ in the space of continuum
$SU(N)$ gauge fields on $T^{2n}$. We restrict $A$ to be in the topologically
trivial sector (otherwise the chiral determinant vanishes). Then,
generically, $\det(\sd_+^A)\ne0$ and we have a map
\be
S^1\to{}S^1\subset{\bf C}\quad,\qquad\theta\mapsto
\det(i\sd_+^{A^{\theta}})/\det(i\sd_+^A)
\label{10}
\ee
(where $S^1\subset{\bf C}$ denotes the unit circle in ${\bf C}$).
The winding number $W_c$ of this map is an obstruction to gauge-invariance
of the chiral determinant (since if the determinant is gauge-invariant then
it is constant around the family $\{A^{\theta}\}_{\theta\in{}S^1}$ and
the winding number is zero). In \cite{AG} the winding number $W_c$ was 
shown to equal the index (which we define here to be
the topological charge of the index bundle)
of a 2-parameter family of Dirac operators
$\sd^{(\theta,t)}$, or equivalently, the index of a Dirac operator $\D_c$
in 2n+2 dimensions, given as follows. (The subscript $c$ here and in the 
following refers to ``continuum''.) 
The family $A^{\theta}$ is extended to a
family $A^{(\theta,t)}=tA^{\theta}$ with $(\theta,t)\in{}B^2$, the
unit disc. ($\{A^{(\theta,t)}\}$ corresponds to a 2-sphere in the orbit
space of gauge fields since the $A^{(\theta,1)}$'s are
all gauge-equivalent.) This determines the family of Dirac operators
$\sd^{(\theta,t)}\equiv\sd^{A^{(\theta,t)}}=\gamma^{\mu}(\partial_{\mu}
+A_{\mu}^{(\theta,t)})$. The Dirac operator $\D_c$ acting on spinor
fields on $B^2\times{}T^{2n}$ is now given by 
\be
\D_c=\Gamma^{\alpha}i(\partial_{\alpha}+\A_{\alpha})\qquad\quad
\alpha=1,\dots,2n+2
\label{12}
\ee
where the gauge field $\A$ on $B^2\times{}T^{2n}$ is given by
$\A_{\mu}(\theta,t,x)=A_{\mu}^{(\theta,t)}(x)\,$, 
$\A_{\alpha}\equiv0$ for $\alpha=2n+1,2n+2$,
and the Dirac matrices in 2n+2 dimensions are chosen as
$\Gamma^{\mu}=\sigma_1\otimes\gamma^{\mu}\,$,
$\Gamma^{2n+1}=\sigma_1\otimes\g5\,$, $\Gamma^{2n+2}=\sigma_2\otimes1$
where $\sigma_j$ $(j=1,2,3)$ are the Pauli matrices and 
$\g5=i^n\gamma^1\gamma^2\cdots\gamma^{2n}$. The $\gamma^{\mu}$'s and
$\sigma_j$'s are taken to be hermitian so $\sd$ is anti-hermitian and
$\D_c$ is hermitian. In (\ref{12}) the derivatives for $\alpha=2n+1,2n+2$
are $\partial_{2n+1}=\frac{\partial}{\partial{}y_1}\,$,
$\partial_{2n+2}=\frac{\partial}{\partial{}y_2}$ where $(y_1,y_2)$ is a
cartesian coordinate system on $B^2$. The polar coordinates $\theta$ and 
$t$ are henceforth viewed as functions of $(y_1,y_2)$. 
Let $\wB^2$ denote another copy of the unit disc, then $B^2\times{}T^{2n}$
and $\wB^2\times{}T^{2n}$ can be glued together along their common boundary
$S^1\times{}T^{2n}$ to get the closed manifold $S^2\times{}T^{2n}$.
An $SU(N)$ vectorbundle over this manifold is defined by taking the
transition function on the common boundary $S^1\times{}T^{2n}$ to be
$\Phi(\theta,x)^{-1}$, where $\Phi(\theta,x)=\phi_{\theta}(x)$.
The topological charge (Pontryargin number) of this bundle is then
$-deg(\Phi)$, i.e. minus the degree (generalised winding number)
of the map $\Phi:S^1\times{}T^{2n}\to{}SU(N)$.
The operator $\D_c$ above extends in a natural way to a Dirac operator 
(also denoted $\D_c$) on the spinor fields in this vectorbundle 
\cite{AG}. $\D_c$ anticommutes with the chirality matrix 
\be
\Gamma_5\,\equiv\,i^{n+1}\Gamma^1\Gamma^1\cdots\Gamma^{2n+2}=\sigma_3\otimes1
\label{19}
\ee
and thus has a well-defined index. The main result of \cite{AG} is that
the obstruction to gauge-invariance discussed above is determined by
this index:
\be
W_c=-\index\D_c
\label{20}
\ee
The index of $\D_c$ can be calculated from the formula
\be
\index\D_c=\Tr(\Gamma_5|_{\ker{}\D_c})
=\Tr(\Gamma_5\,e^{-\tau\D_c^2})\quad\forall\tau>0
\label{22}
\ee
This can be calculated in the $\tau\to0$ limit by familiar
techniques, leading to \cite{AG}
\be
\index\D_c=-deg(\Phi)
\label{23}
\ee
Thus the obstruction is determined to be $W_c=deg(\Phi)$.

We now put a hypercubic lattice on $T^{2n}$, with lattice spacing $a$, and
proceed to describe a lattice version of the preceding.
Chiral gauge theory can be formulated on the lattice in the overlap
formalism \cite{overlap}. This approach can be reformulated as a functional
integral approach with lattice Dirac operator $D$ satisfying the GW
relation (\ref{1}) and $\g5-$hermiticity condition (\ref{7a}) \cite{L1,L2}. 
Let $\C$ denote the finite-dimensional space of lattice spinor fields 
on $T^{2n}$. The chiral projections $P_{\pm}=\frac{1}{2}(1\pm\g5)$ and
$\hP_{\pm}=\frac{1}{2}(1\pm\hg5)$ determine decompositions 
$\C=\C_+\oplus\C_-$ and $\C=\hC_+\oplus\hC_-$ respectively.
The (right-handed) lattice chiral determinant in this setting is
\be
\det(iD_+^U)=\la{}v_-\,,\hw_+(U)\ra
\label{29}
\ee
where $v_-$ and $\hw_+$ are
unit volume elements on $\C_-$ and $\hC_+$ respectively. These are unique
up to phase factors; they can be written as
$v_-=v_1\wedge\cdots\wedge{}v_d$ and $\hw_+=\hw_1\wedge\cdots\hw_d$
where $v_1,\dots{}v_d$ and $\hw_1,\dots,\hw_d$ are orthonormal bases 
for $C_-$ and $\hC_+$ respectively. 
$v_-$ and $\hw_+$ are the many-body groundstates in the overlap
formulation \cite{overlap}, and correspond to the chiral fermion measures
in the formulation of ref.'s \cite{L1,L2}. 
Note that $\hg5=\g5(1-aD^U)$ depends on the lattice gauge field
$U$, so the subspace
$\hC_+$ and volume element $\hw_+$ likewise depend on $U$. On the other
hand, since the usual chiral decomposition $\C=\C_+\oplus\C_-$ does not
involve $U$, neither does $v_-$.
We are assuming $\dim\C_{\pm}=\dim\hC_{\pm}\equiv{}d$
(otherwise the chiral determinant vanishes). This is equivalent to assuming
$\index{}D^U=0\,$, i.e. $U$ is in the 
topologically trivial sector \cite{overlap,L2}.
The space of lattice gauge fields will typically contain a subset of measure
zero where $D^U$ is not defined. In the case of the overlap Dirac operator
such fields can be excluded by imposing a condition of the form
$||1-U(p)||<\epsilon$ on the plaquette products of $U$ \cite{L(local)}.
This condition is automatically satisfied close to the classical continuum
limit since $1-U(p)=a^2F_{\mu\nu}(x)+O(a^3)$. We will assume that the same 
is true for the general $D$ that we are considering here.

Let $\{\phi_{\theta}\}_{\theta\in{}S^1}$ be a family of $SU(N)$ lattice
gauge transformations, then the winding number $W$ of the map
\be
S^1\to{}S^1\quad,\quad\theta\mapsto
\la{}v_-\,,\hw_+(\phi_{\theta}\cdot{}U)\ra/\la{}v_-\,,w_+(U)\ra
\label{30}
\ee
is an obstruction to gauge-invariance of the chiral determinant (\ref{29})
just as in the continuum setting. This was recently studied in \cite{DA}
where $W$ was shown to reduce to $W_c$ in the classical continuum limit.
In the following we will show that this
obstruction is related to the index of a family of lattice Dirac operators,
defined as the index of a Dirac operator $\D$ in 2n+2 dimensions, in
complete analogy with the continuum relation (\ref{20}).

The action of $\phi_{\theta}$ on $U$ generates a 
circle-family $\{\phi_{\theta}\cdot{}U\}_{\theta\in{}S^1}$ in the 
space $\U$ of lattice gauge fields. Choose a disc-family 
$\B^2=\{U^{(\theta,t)}\}_{(\theta,t)\in{}B^2}$ in $\U$ 
such that $U^{(\theta,1)}=\phi_{\theta}\cdot{}U$. 
(Such a family might not exist in general due to the restrictions on
$\U$ needed to ensure that $D$ is well-defined. However its existence
is guaranteed close to the classical continuum limit: we can take
the lattice transcript of the continuum family $A^{(\theta,t)}$.)
This determines a family of lattice Dirac operators 
$D^{(\theta,t)}=D^{U^{(\theta,t)}}$. 
Setting $\hG^1=\sigma_1\otimes\hg5$
(where $\hg5=\hg5^{(\theta,t)}$ is given by (\ref{7}) with $D=D^{(\theta,t)}$)
and $\hG^2=\Gamma^2=\sigma_2\otimes1$, we define the Dirac operator in
2n+2 dimensions in the lattice setting to be
\be
\D=\hG^{\alpha}i(\partial_{\alpha}+A_{\alpha})\qquad\quad\alpha=1,2
\label{31}
\ee
The derivatives are with respect to
the continuous cartesian coordinates $(y_1,y_2)$ on $B^2$ and we have
introduced a continuum SU(N) gauge field $A=A_{\alpha}dy_{\alpha}$
on $B^2$ with $A_{\alpha}(y_1,y_2,x)$ a function of lattice
site $x$ as well as $(y_1,y_2)$. 
$\D$ extends in a natural way to an elliptic 1st order differential
operator on the vectorfields
with values in a vectorbundle over the closed manifold
$S^2=B^2\cup_{S^1}\wB^2$ as follows. The fibre of the vectorbundle is
${\bf C}^2\otimes\C$ (i.e. the representation space of the Pauli matrices
tensored with the finite-dimensional vectorspace of lattice spinor fields
on $T^{2n}$) and the transition function at the common boundary $S^1$
of $B^2$ and $\wB^2$ is $1\otimes\Phi^{-1}$ where 
$\Phi(\theta)=\phi_{\theta}$. A vectorfield in this vectorbundle 
consists of a function $\Psi(\theta,t)$ on $B^2$ together with a  
function $\wPsi(\theta,s)$ on $\wB^2$, both taking values in 
${\bf C}^2\otimes\C$, and related at the common boundary $S^1$ by
\be
\wPsi(\theta,1)=\Phi(\theta)^{-1}\cdot\Psi(\theta,1)\,\equiv\,
1\otimes\phi_{\theta}^{-1}\cdot\Psi(\theta,1)
\label{32}
\ee
$\D$ is defined on $\Psi$ by (\ref{31}), and is defined to act on $\wPsi$ as
$\hG_U^{\alpha}i(\partial_{\alpha}+\wA_{\alpha})$
where $\hG_U^1=\sigma_1\otimes\hg5^U$ and $\hG_U^2=\Gamma^2$.
The gauge-covariance
of $D$ implies that $D^{(\theta,1)}=D^{\phi_{\theta}\cdot{}U}=
\phi_{\theta}\circ{}D^U\circ\phi_{\theta}^{-1}$ and 
$\hg5^{(\theta,1)}=\phi_{\theta}\circ\hg5^U\circ\phi_{\theta}^{-1}$.
Using these it is easily checked that $\D$ respects the relation (\ref{32}),
and is therefore a well-defined operator on the vectorfields in the above 
vectorbundle over $S^2$, provided the gauge field 
$\wA=\wA_{\alpha}dy_{\alpha}$ on $\wB^2$ is related to the field $A$ on
$B^2$ at the common boundary $S^1$ by
\be
\wA(\theta,1,x)=\phi_{\theta}(x)^{-1}A(\theta,1,x)\phi_{\theta}(x)
+\phi_{\theta}(x)^{-1}d_{\theta}\phi_{\theta}(x)
\label{31b}
\ee
(E.g. we can take $A\equiv0$ and 
$\wA(\theta,s,x)=s\phi_{\theta}(x)^{-1}d_{\theta}\phi_{\theta}(x)$
in terms of polar coordinates $(\theta,s)$ on $\wB^2$.)

The space of vectorfields in the above vectorbundle over $S^2$ is denoted by
$\V$ in the following. The chirality operator $\Gamma_5=\sigma_3\otimes1$
determines a chiral decomposition $\V=\V_+\oplus\V_-$.
The ellipticity of $\D$ follows easily from the
facts that $\sigma_1\otimes\hg5$ anticommutes with $\sigma_2\otimes1$
and $(\sigma_1\otimes\hg5)^2=(\sigma_2\otimes1)^2=1\otimes1$. 
Also, $\D$ is formally self-adjoint with respect to the natural inner
product in $\V$ since $\sigma_1\otimes\hg5$ and $\sigma_2\otimes1$ are
self-adjoint on ${\bf C}^2\otimes\C$.
$\D$ anticommutes with $\Gamma_5=\sigma_3\otimes1$ and therefore
has a chiral decomposition $\Bigl({0 \atop \D_+}\;{\D_- \atop 0}\Bigr)$
and $\index\D=\dim\,\ker\D_+-\dim\,\ker\D_-$. The following formula for the 
index is derived below:

\noindent {\it Theorem.}
\be
\index\D=-\frac{1}{2\pi{}i}\left(\,\int_{\B^2}\Tr(\hP_+d\hP_+d\hP_+)
+\frac{1}{2}\int_{{\cal S}^1}\Tr(\phi_{\theta}^{-1}d_{\theta}\phi_{\theta}
\hg5^U)\right)
\label{100}
\ee
Here $\hP_+$ is to be viewed as a function on the space $\U$ of lattice
gauge fields whose values are operators on $\C$ (i.e. finite-dimensional
matrices), and $d$ is the exterior derivative on $\U$. 
Thus the first integrand
is a 2-form on $\U$ and can be integrated over the disc $\B^2$ in $\U$
to get a ${\bf C}$-number.
The second integrand is a 1-form on the boundary ${\cal S}^1$ of $\B^2$,
with $\hg5^U=\hg5^{(0,1)}$ constant. 

By Eq.(3.11) of \cite{DA} the obstruction (winding number) $W$ 
associated with the map (\ref{30}) equals the right-hand side 
of (\ref{100}) without the minus sign. It follows that 
\be
W=-\index\D\,.
\label{101}
\ee
This is the promised lattice analogue of (\ref{20}). Since $W$ reduces to 
$W_c=deg(\Phi)$ in the classical continuum limit \cite{DA},  
it follows that $\index\D$ reduces to $\index\D_c=-deg(\Phi)$ in
this limit. 
The 2-form in the first term in 
the right-hand side of (\ref{100}) has appeared previously
in the overlap formalism in \cite{Neu(PRD)}, where it was interpreted
as a form of Berry's curvature. (The Berry phase is associated with the
state $w_+(U)$ in (\ref{29}).) It is interesting to note that a version of 
this 2-form also arises in the context of the quantised Hall effect
\cite{Seiler}. The second term in (\ref{100}) arises in \cite{DA} as the
integral of the covariant gauge anomaly (and vanishes in the special case
where $U=1$).

The index formula (\ref{100}), together with (\ref{a9}) for the index 
density, and the relation (\ref{101}) are the main results of this paper.
The proof is as follows. Analogously to (\ref{22}) we have
\be
\index\D=\Tr(\Gamma_5\,e^{-\tau\D^2})\quad\forall\tau>0\,.
\label{a1}
\ee
This can be evaluated in the $\tau\to0$ limit
by the same familiar techniques used to evaluate (\ref{22}) in \cite{AG}:
It is seen to be the sum of
a contribution from the $B^2$ part of $S^2=B^2\cup{}\wB^2$,
given by
\be 
\int_{B^2}d^2y\,a^4\sum_xq_{\D}(x,y_1,y_2)
\label{a2}
\ee
where
\be
q_{\D}(x,y_1,y_2)=\lim_{\tau\to0}\ \int_{-\infty}^{\infty}
\frac{d^2k}{(2\pi)^2}\,\tr(\Gamma_5\,e^{-ik\cdot{}y}
(e^{-\tau\D^2})e^{ik\cdot{}y})(x,x)\,,
\label{a3}
\ee
and an analogous contribution from the $\wB^2$ part. 
In (\ref{a3}) the trace is over spinor and flavour indices;
${\cal O}(x,y)$ denotes the kernel function of an operator ${\cal O}$ 
on scalar lattice fields. 
Setting $\nabla_{\alpha}=\partial_{\alpha}+A_{\alpha}$ we find from
(\ref{31}) that
\be
-\D^2&=&\hG^{\alpha}\nabla_{\alpha}\hG^{\beta}\nabla_{\beta}
=\hG^{\alpha}\hG^{\beta}\nabla_{\alpha}\nabla_{\beta}
+\hG^{\alpha}\lb\nabla_{\alpha}\,,\hG^1\rb\nabla_1 
=\nabla_{\alpha}\nabla_{\alpha}+i\sigma_3\hg5{}F_{12}
+(\hg5\nabla_1\hg5-i\sigma_3\nabla_2\hg5)\nabla_1 \nonumber \\
&=&\partial^2+(2A_1+\hg5\nabla_1\hg5-i\sigma_3\nabla_2\hg5)\partial_1
+2A_2\partial_2+\nabla_{\alpha}A_{\alpha}+i\sigma_3\hg5{}F_{12}
+(\hg5\nabla_1\hg5-i\sigma_3\nabla_2\hg5)A_1
\label{a5}
\ee
where $\nabla_{\alpha}\hg5\,\equiv\,\partial_{\alpha}\hg5
+\lb{}A_{\alpha},\hg5\rb$ (as in \cite{L2}); for notational simplicity
we have omitted the $\otimes$ symbol.
After substituting this in (\ref{a2}) and making a change of variables 
$k_j\to\tau^{-1/2}k_j$ we find
\be
q_{\D}(x,y_1,y_2)&=&\lim_{\tau\to0}\ \frac{1}{\tau}
\int_{-\infty}^{\infty}\frac{d^2k}{(2\pi)^2}\,\tr(\sigma_3\,\exp
\{-k^2+\sqrt{\tau}(2A_1+\hg5\nabla_1\hg5-i\sigma_3\nabla_2\hg5)ik_1
+\sqrt{\tau}(2A_2)ik_2 \nonumber \\
& &\qquad\qquad\qquad\qquad\qquad\qquad\quad
+\;\tau(\nabla_{\alpha}A_{\alpha}+\hg5\nabla_1\hg5+
i\sigma_3(\hg5{}F_{12}-\nabla_2\hg5{}A_1))\})(x,x)
\label{a6}
\ee
This can be calculated by expanding the integrand in powers of
$\sqrt{\tau}$. 
Terms with odd powers of $\sigma_3$ give vanishing contribution, as do
terms with $k$-dependence of the form
$e^{-k^2}k_1^pk_2^q$ where either $p$ or
$q$ is odd. The remaining terms are
\be
\tau{}ie^{-k^2}\tr((2A_1\nabla_2\hg5+\hg5\nabla_1\hg5\nabla_2\hg5)k_1^2
+\hg5{}F_{12}-\nabla_2\hg5{}A_1)(x,x)+O(\tau^{3/2})
\label{a7}
\ee
where we have used the fact that $\hg5\nabla_{\alpha}\hg5
=-\nabla_{\alpha}\hg5\hg5$
(an easy consequence of $\hg5^2=1$). Evaluating the integral in (\ref{a6})
with this integrand, we find that the contributions from the 
$2A_1\nabla_2\hg5{}k_1^2$ and $-\nabla_2\hg5{}A_1$ terms in $\tr(\cdots)$
in (\ref{a7}) cancel, resulting in
\be
q_{\D}(x,y_1,y_2)=\frac{-1}{2\pi{}i}\Bigl(\,\frac{1}{4}\tr(\hg5\nabla_1
\hg5\nabla_2\hg5)(x,x)+\frac{1}{2}\tr(\hg5{}F_{12})(x,x)\Bigr)
\label{a9}
\ee
Modulo the numerical factor $-1/2\pi$, this coincides with L\"uscher's
topological field \cite{L2}
\be
q(x,y_1,y_2)=-i\tr({\textstyle \frac{1}{4}}
\hg5\lb\nabla_1\hP_-,\nabla_2\hP_-\rb+{\textstyle \frac{1}{4}}
\lb\nabla_1\hP_-,\nabla_2\hP_-\rb\hg5+{\textstyle \frac{1}{2}}F_{12}\hg5)
(x,x)\,.
\label{a10}
\ee
(Note that $\hg5\lb\nabla_1\hP_-,\nabla_2\hP_-\rb=
\lb\nabla_1\hP_-,\nabla_2\hP_-\rb\hg5=\frac{1}{2}\hg5\nabla_1
\hg5\nabla_2\hg5$ (an easy consequence of $\hg5\nabla_{\alpha}\hg5
=-\nabla_{\alpha}\hg5\hg5$) and that
$\tr(\hg5(x,x)F_{12}(x))=\tr(F_{12}(x)\hg5(x,x))$.)
Summing (\ref{a9})
over the lattice sites gives (cf. Appendix B of \cite{L2})
\be
a^4\sum_xq_{\D}(x,y_1,y_2)=\frac{-1}{2\pi{}i}\Tr({\textstyle \frac{1}{4}}
(\hg5\partial_1\hg5\partial_2\hg5)-{\textstyle \frac{1}{2}}
\partial_1(A_2\hg5)+{\textstyle \frac{1}{2}}\partial_2(A_1\hg5))
\label{a11}
\ee
The contribution to (\ref{a2}) from the first term in (\ref{a11})
gives the first term in the index formula (\ref{100}). The contribution
to (\ref{a2}) from the remaining terms in (\ref{a11}) reduces to 
$-\frac{1}{2}\int_{S^1}\Tr(A(\theta,1)\hg5^{(\theta,1)})$ 
in polar coordinates. 
The analogous contribution to $\index\D$ from the $\wB^2$ part is only
$+\frac{1}{2}\int_{S^1}\Tr(\wA(\theta,1)\hg5^U)$ (since $\hg5^U$ is constant). 
Adding these and using (\ref{31b}) we get the second term in (\ref{100}).

I thank Prof.'s Ting-Wai Chiu, Kazuo Fujikawa, Martin L\"uscher and Herbert
Neuberger for interesting discussions/correspondence.
The hospitality and financial support of the National
Centre for Theoretical Sciences in Taiwan, where this work was begun,
is gratefully acknowledged. The author is supported
by an ARC postdoctoral fellowship.


\begin{thebibliography}{XXX}

\bibitem{AS(single)}
M. F. Atiyah and I. M. Singer, Ann. Math. {\bf 87}, 484 (1968)

\bibitem{AS(families)}
M. F. Atiyah and I. M. Singer, Ann. Math. {\bf 93}, 119 (1971)

\bibitem{Fuji(cont)}
K. Fujikawa, Phys. Rev. Lett {\bf 42}, 1195 (1979); 
Phys. Rev. D {\bf 21}, 2848 (1980).

\bibitem{tHooft}
G. 't Hooft, Phys. Rev. Lett. {\bf 37} (1976) 8; Phys. Rev. D {\bf 14}
(1976) 3432; R. Jackiw and C. Rebbi, Phys. Rev. Lett. {\bf 37} (1976) 172;
C. Callan, R. Dashen and D. Gross, Phys. Lett. B {\bf 63} (1976) 334;
Phys. Rev. D {\bf 17} (1978) 2717

\bibitem{Schwarz}
A. S. Schwarz, Comm. Math. Phys. {\bf 64}, 233 (1979)

\bibitem{AG}
L. Alvarez-Gaum\'e and P. Ginsparg, Nucl. Phys. B {\bf 243}, 449 (1984)

\bibitem{AS(anomaly)}
M. F. Atiyah and I. M. Singer, Proc. Natl. Acad. Sci. USA {\bf 81},
2597 (1984)

\bibitem{Nash(book)}
For an overview and further references, see, e.g.,
C. Nash, {\it Differential topology and quantum field theory}
(Academic Press, London, 1991)

\bibitem{Wilson}
K. Wilson, Phys. Rev. D {\bf 14}, 2445 (1974)

\bibitem{NN}
H. B. Nielsen and M. Ninomiya, Nucl. Phys. B {\bf 185}, 20 (1981).

\bibitem{Smit}
J. Smit and J. Vink, Nucl. Phys. B {\bf 286}, 485 (1987)

\bibitem{overlap}
R. Narayanan and H. Neuberger, Phys. Lett B {\bf 302}, 62 (1993);
Nucl. Phys. B {\bf 412}, 574 (1994); {\bf 443}, 305 (1995)

\bibitem{Neu(PLB1+2)}
H. Neuberger, Phys. Lett. B {\bf 417}, 141 (1998); {\bf 427}, 353 (1998).

\bibitem{Hasenfratz}
P. Hasenfratz, in {\it Nonperturbative quantum field physics} (Pe\~niscola
1997), eds. M. Asorey, A. Dobado (World Scientific, Singapore, 1998)

\bibitem{Laliena}
P. Hasenfratz, V. Laliena and F. Niedermayer, Phys. Lett. B {\bf 427},
125 (1998).

\bibitem{L(PLB)}
M. L\"uscher, Phys. Lett. B {\bf 428}, 342 (1998)

\bibitem{GW}
P. Ginsparg and K. G. Wilson, Phys. Rev. D {\bf 25}, 2649 (1982).

\bibitem{Fuji(NPB)}
K. Fujikawa, Nucl. Phys. B {\bf 546}, 480 (1999)

\bibitem{DA(Chiral99)}
D. H. Adams, hep-lat/0001014

\bibitem{KY+DA+Suzuki+ChiuH}
Y. Kikukawa and A. Yamada, Phys. Lett. B {\bf 448}, 265 (1999);
D. H. Adams, hep-lat/9812003;
H. Suzuki, Prog. Theor. Phys. {\bf 102}, 141 (1999);
T.-W. Chiu and T.-H. Hsieh, hep-lat/9901011

\bibitem{DA(inprogress)}
D. H. Adams, work in progress

\bibitem{L1}
M. L\"uscher, Nucl. Phys. B {\bf 549}, 295 (1999)

\bibitem{L2}
M. L\"uscher, Nucl. Phys. B {\bf 568}, 162 (2000)

\bibitem{DA}
D. H. Adams, Nucl. Phys. B {\bf 589}, 633 (2000)

\bibitem{L(local)}
P. Hern\'andez, K. Jansen and M. L\"uscher, Nucl. Phys. B 
{\bf 552}, 363 (1999)

\bibitem{Neu(PRD)}
H. Neuberger, Phys. Rev. D {\bf 59}, 085006 (1999)

\bibitem{Seiler}
J. Avron, R. Seiler and B. Simon, Phys. Rev. Lett. {\bf 51}, 51 (1983)


\end{thebibliography}
\end{document}